# Nanoantenna–Microcavity Hybrids with Highly Cooperative Plasmonic–Photonic Coupling

*Jui-Nung Liu,[†] Qinglan Huang,[†] Keng-Ku Liu,[‡] Srikanth Singamaneni,[‡,]\**

*and Brian T. Cunningham[†,]\**

**Affiliations**

[†]Department of Electrical and Computer Engineering, Department of Bioengineering, Micro and Nanotechnology Laboratory, University of Illinois at Urbana-Champaign, Urbana, Illinois 61801, United States

[‡]Department of Mechanical Engineering and Materials Science, Institute of Materials Science and Engineering, Washington University in St. Louis, St. Louis, Missouri 63130, United States

*Corresponding authors.  B.T.C.: bcunning@illinois.edu;  S.S.: singamaneni@wustl.edu

**Abstract**

Nanoantennas offer the ultimate spatial control over light by concentrating optical energy well below the diffraction limit, whereas their quality factor ($Q$) is constrained by large radiative and dissipative losses. Dielectric microcavities, on the other hand, are capable of generating a high $Q$-factor through an extended photon storage time, but have a diffraction-limited optical mode volume. Here we bridge the two worlds, by studying an exemplary hybrid system integrating plasmonic gold nanorods acting as nanoantennas with an on-resonance dielectric photonic crystal (PC) slab acting as a low-loss microcavity, and, more importantly, by synergistically combining their advantages to produce a much stronger local field enhancement than that of the separate entities. To achieve this synergy between the two polar opposite types of nanophotonic resonant elements, we show that it is crucial to coordinate both the dissipative loss of the nanoantenna and the $Q$-factor of the low-loss cavity. In comparison to the antenna–cavity coupling approach using a Fabry–Perot resonator, which has proved successful for resonant amplification of the antenna's local field intensity, we theoretically and experimentally show that coupling to a modest-$Q$ PC guided resonance can produce a greater amplification by at least an order of magnitude. The synergistic nanoantenna–microcavity hybrid strategy opens new opportunities for further enhancing nanoscale light–matter interactions to benefit numerous areas such as nonlinear optics, nanolasers, plasmonic hot carrier technology, and surface-enhanced Raman and infrared absorption spectroscopies.

**Main Text**

Optical nanoantennas are key elements for the subwavelength manipulation and concentration of light.[1-3] The intense, strongly localized electromagnetic field enhancement created by nanoantennas is at the heart of a rich variety of technologies and applications, including surface-enhanced Raman scattering (SERS),[4] surface-enhanced infrared absorption (SEIRA),[5-7] plasmon-induced photodetection,[8] spontaneous emission enhancement,[9] photothermal biosensing,[10] nonlinear nanophotonics,[11] and nanolasing.[12] To maximize the local field enhancement, various methods have been studied including, for example, changing the antenna's dimension and loading materials to tune resonance,[6, 13] using a lower-loss material,[14, 15] impedance matching between the input excitation and the nanoantenna,[16] and Fano resonance engineering.[17, 18] In addition to these techniques, reducing the mode volume of the antenna by shrinking the gap size has been extensively employed to effectively increase the enhancement.[19, 20] When entering the sub-nanometer gap regime, however, it has been recently shown that quantum mechanical effects such as nonlocality and electron tunneling stop the enhancement factor from further increasing monotonically.[21, 22] Thus, to further boost the field enhancement of the nanoantenna, a new strategy is needed.

Integrating nanoantennas with oscillating photonic building blocks, such as an evanescent diffraction order,[6, 23, 24] a plasmonic crystal,[15] or a Fabry–Perot (FP) cavity,[7, 25, 26] can merge the advantages of both the high spatial mode energy density (ultrasmall mode volume) of the nanoantenna and the large spectral mode energy density (long photon storage time) of the resonant cavity, which thus provides a new degree of freedom to enhance the nanoantenna resonance. The nanoantenna–cavity hybrid approach has been shown to further increase the local field intensity of the nanoantenna by up to one order of magnitude, and therefore can enhance light–matter interactions to a greater extent than the separate entities alone. Accordingly, to further improve enhancement in the hybrid approach, it is intuitive to employ a high-quality-factor (high-$Q$) dielectric optical microcavity[27] in order to benefit from a stronger optical feedback.[28] When matching the high-$Q$ cavity mode to the antenna resonance for the purpose of exploiting their combined advantages, however, it has been shown that either the antenna resonance was strongly suppressed by the destructive Fano interference, or the microcavity mode was spoiled by the intrinsic ohmic loss in the nanoantenna,[29-33] implying that a low cooperativity between the two distinct elements was obtained and their combined advantages cannot be exploited simultaneously. Achieving synergistic nanoantenna–microcavity coupling necessary for greater resonant amplification of local fields, therefore, remains a challenge.

In this work, we tackle this challenge by studying a hybrid system combining the simplest type of the nanoantenna without loss of generality, gold nanorods (AuNRs),[34] and a resonant dielectric photonic crystal (PC) slab[35, 36] acting as an open microcavity. Here, the PC slab is designed to possess two types of cavity resonances: a broadband FP cavity and a narrowband PC guided resonance (PCGR), which can be activated under specific illumination conditions to pair and hybridize with the AuNR resonance. The platform therefore enables a direct comparison between the two types of cavity-coupling approaches for local field enhancement. We experimentally demonstrate, through observations of SERS, that coupling to the PCGR provides an at least one order of magnitude greater amplification of the antenna's local field intensity compared to that provided by coupling to the FP cavity, in agreement with our simulated predictions. To achieve the highly cooperative, PCGR-powered resonant amplification of local fields, further analysis using temporal coupled-mode theory (TCMT) reveals that it is essential to have a judicious combination of the absorption quality factor ($Q_{abs}$) and the radiation quality factor ($Q_{rad}$) of the hybrid system. Finally, we develop strategies to guide future efforts in this direction for achieving ultrastrong hybrid near-field enhancement.

We have performed full-wave electromagnetic simulations (described in Supporting Information) to study near- and far-field optical properties of the hybrid system. Figure 1 schematically illustrates the concept of the synergistic nanoantenna–cavity hybrid and the unit cell of the simulation. Bulk dielectric substrates with a high refractive index such as $TiO_2$ have been shown to shift resonances of the plasmonic nanoparticle antenna.[37] When the dielectric substrate is equipped with an oscillating photonic mode that can resonantly interact with the nanoantenna, the hybrid system exhibits new and interesting physics. The wavelength-scale-thick $TiO_2/SiO_2$ films located on Si act as a FP cavity in the $z$ direction, producing broadband ripples in the reflection spectrum by thin-film interference.[38] On the other hand, when meeting the phase-matching condition, the transverse-magnetic (TM, with an incident electric field $\mathbf{E}_{inc}$ on the $x$-$z$ plane) polarized incident plane wave can excite the delocalized, long-lifetime PCGR that is supported by the corrugated high-index $TiO_2$ waveguiding layer.[36] Excitation of the TM-polarized PCGR creates a sharp peak in the far-field reflection spectrum, and generates enhanced near fields on the PC surface pointing in the $x$ and $z$ directions.[39] In comparison to a solitary AuNR (with a resonance wavelength $\lambda_{ant}$ and a quality factor $Q_{ant}$) without a cavity, integrating the AuNR with an on-resonance PCGR (with a resonance wavelength $\lambda_{PCGR} \approx \lambda_{ant}$ and a quality factor $Q_{PCGR}$ for the bare PCGR) can reshape near-field optical properties of the AuNR via the following processes:

(i) The PCGR mode effectively captures directional plane-wave input energy, and then feeds the AuNR with a much stronger resonant excitation through near-field coupling (illustrated in Figure 1). On the other hand, a solitary AuNR is directly excited by the plane-wave input with a lower efficiency, where AuNR's radiation pattern weakly matches the directional excitation signal.[40]

(ii) Near-field coupling to the PCGR mode helps collect the electromagnetic energy scattered from the AuNR, and circulates it inside the low-loss cavity. In contrast, the energy of a solitary AuNR is scattered away without being recycled.

With these benefits, the hybrid system can therefore strongly amplify local fields of the AuNR, with a cavity-inherited hybrid resonance $Q$-factor ($Q_{hyb}$) higher than $Q_{ant}$, as illustrated in the top left inset of Figure 1.

The AuNRs in the simulation sit at the TiO$_2$ groove center, and have a surface density of 1 /μm$^2$, which is low enough to avoid near-field plasmonic coupling between adjacent AuNRs.[41] When AuNRs are oriented in the $x$ direction (therefore aligned with the near-field polarization of the on-resonance PCGR), efficient intermodal coupling between the AuNR and the PCGR is enabled. The simulated far-field reflectance spectrum at normal incidence (Figure 2A) shows that peak reflectance ($R_{peak}$) of the hybrid drops compared to that of the bare PC slab. In addition to the far-field behavior, the antenna–PCGR coupling also profoundly modifies the near-field characteristics. We computed the average near-field intensity $\langle|\mathbf{E}|^2\rangle$ right on the AuNR surface $\Sigma_{AuNR}$ ($\langle|\mathbf{E}|^2\rangle \equiv \int_{\Sigma_{AuNR}} |\mathbf{E}|^2 d^2\mathbf{r} \Big/ \int_{\Sigma_{AuNR}} d^2\mathbf{r}$, where the near-field strength $|\mathbf{E}|$ is normalized to the incident electric-field strength $|\mathbf{E}_{inc}|$) for the hybrid system, the solitary AuNR on bulk TiO$_2$, and the bare PC slab (calculated over the same surface $\Sigma_{AuNR}$ but in the absence of the physical AuNR), respectively (Figure 2B). The solitary AuNR on the bulk TiO$_2$ dielectric exhibits an apparent dipolar antenna resonance at $\lambda_{ant}$ = 633 nm. The near-field $\langle|\mathbf{E}|^2\rangle$ spectra unambiguously show that the integration of the two on-resonance constituents leads to a local field intensity approximately

two orders of magnitude greater than that of the solitary AuNR or bare PC slab, with a resonance linewidth being considerably narrowed at the same time compared to that of the solitary AuNR ($Q_{ant}$ = 15 and $Q_{hyb}$ = 184). The Q-factor of the hybrid $Q_{hyb}$ is approximately one half of that of the bare cavity ($Q_{PCGR}$ = 435) due to efficient resonant hybrid coupling, and the mechanism behind this observation will be further elucidated in this report.

The advantages of the hybrid integration can be more clearly demonstrated by spatial profiles of the near fields at resonance (Figure 2, C and E). As featured by the standing-wave pattern along the *x* direction around the $TiO_2$ slab formed by interference of counter-propagating waveguide modes (Figure 2E), the phase-matching incident plane wave efficiently excites the PCGR. The PCGR overlaps the AuNR's mode both *spatially* and *spectrally*, which slightly lowers the field intensity of the PCGR mode compared to that of the bare cavity (Figure S1B, left panel) due to introduced ohmic loss but, in return, strongly modifies and amplifies the AuNR's local field intensity. Also, the optical energy is tightly concentrated around the AuNR, showing that the hybrid system retains the nanoantenna's capability for subwavelength light concentration. The PCGR-excited antenna also partially redirects its energy into the FP cavity underneath,[25] leading to a standing-wave pattern along the *z* direction in $SiO_2$ (Figure 2E) and a slightly larger (1.8×) peak $\langle|\mathbf{E}|^2\rangle$ value compared to that of the hybrid without an underlying FP cavity (Figure S2). The spectral and spatial near-field profiles of the hybrid system here (Figure 2, B, C, and E) demonstrate a synergistic nanoantenna–cavity interaction that is fundamentally different from the behaviors of the previously reported hybrid systems with high-Q microcavities, where either the antenna or the cavity resonance is strongly suppressed/spoiled when the two components are on-resonance.[29-33]

On the other hand, when the AuNR is *y*-oriented, the antenna–cavity coupling is prohibitively weak due to mismatch of the polarizations. Thus, $R_{\text{peak}}$ of the hybrid system is almost the same as that of the bare PC slab (Figure 2A). In addition to the far-field response, this low coupling efficiency is also depicted by the near-field properties—a low peak in the $\langle|\mathbf{E}|^2\rangle$ spectrum with $Q_{\text{hyb}} \approx Q_{\text{PCGR}}$ (Figure 2B) and the spatial field profiles showing almost no interactions between the two constituents (Figure 2, D and F). Compared to the orientation, the hybrid coupling is weakly dependent on the location of the AuNR (Figure S3).

In the experiment, the PC slab was fabricated using a top-down approach based on deep ultraviolet (DUV) lithography, and AuNRs were chemically synthesized from the bottom up[42] (Supporting Information). The synthesized AuNRs on the flat TiO$_2$-coated (thickness ≈ 25 nm, to mimic the local dielectric environment of the AuNR) glass substrate exhibit a measured extinction peak at 633 nm (Figure 2H), which closely matches the simulation. The AuNRs were dispersed on the PC slab, with a surface density (~1.5 /μm$^2$) close to the simulation (Figure 2G). Far-field reflectance spectra of the PC slab with and without AuNRs were measured under TM-polarized collimated light (Figure 2H), clearly showing a reduction of $R_{\text{peak}}$ after incorporating AuNRs, which agrees with the simulated predictions. Because the illumination spot (area ≈ 5 mm$^2$) contains multiple AuNRs that are oriented randomly on the *x-y* plane, the reduction of $R_{\text{peak}}$ in the experiment is between the simulated reduction values of hybrids with either *x*-oriented or *y*-oriented AuNRs. Besides the far-field characteristics, in a later section we will experimentally probe the numerically predicted strong amplification of local fields — the focus of this report — with surface-enhanced spectroscopy.

The synergistic hybrid enhancement also exhibits continuous spectral tunability within the antenna's resonance bandwidth. The experimentally measured far-field reflectance spectrum of the hybrid is shown in Figure 3A as a function of incident angle $\theta_{inc}$. Here, we focus on the long-wavelength-branch resonance in order to observe the behavior of the hybrid while the PCGR is gradually detuned from the antenna's resonance wavelength $\lambda_{ant}$. The peak wavelengths of the experimental far-field spectra agree well with the analytical dispersion of the PCGR determined based on the phase-matching condition [$\beta = (2\pi/\lambda)\sin\theta_{inc} \pm 2\pi/P$, where $\lambda$ is the wavelength in free space, $P$ is the period of the PC slab, and $\beta$ is the propagation constant of the TM$_0$ waveguide mode supported by the air/TiO$_2$/SiO$_2$ slab with a slab thickness of $t_{TiO2}$].[39] Moreover, the peak wavelengths of the simulated near-field $\langle|\mathbf{E}|^2\rangle$ spectra of the hybrid (with $x$-oriented AuNRs) closely match the analytical dispersion curve of the PCGR. This shows that (i) the near- and far-field optical properties of the hybrid are highly correlated, and (ii) the PCGR-powered hybrid enhancement is wavelength-tunable by varying $\theta_{inc}$ (also see Figure S4).

The corresponding $Q_{PCGR}$, $Q_{hyb}$, and peak $\langle|\mathbf{E}|^2\rangle$ value of the hybrid were evaluated as a function of $\theta_{inc}$ with simulation (Figure 3B). When the PCGR deviates from the antenna resonance (but still within the antenna's bandwidth, $\theta_{inc} \leq 8°$), interestingly, the hybrid still retains a high enhancement. The behavior exhibits a slightly different type of hybrid cooperation—a weaker antenna resonance is compensated by a stronger cavity resonance as depicted by a higher $Q_{PCGR}$ (Figure 3B). As the PCGR is further detuned beyond the antenna's bandwidth ($\theta_{inc} \geq 10°$), the growth rate of $Q_{PCGR}$ slows down, and the antenna resonance becomes weaker, therefore leading to a reduced peak $\langle|\mathbf{E}|^2\rangle$ value in this regime. Due to the large difference between $Q_{ant}$ and $Q_{PCGR}$, the antenna resonance can easily encompass the narrowband, spectrally tunable PCGR within its broad resonance bandwidth, which relaxes the wavelength tolerance and increases the degree of robustness and tunability for resonant hybrid enhancement.

The wavelength-tunable property allows the narrow-linewidth, synergistic hybrid near-field enhancement to match a specific wavelength of interest in applications, for example, a pump frequency in resonantly enhanced second- and third-harmonic generation,[11] a radiative transition energy of the quantum emitter for spontaneous emission enhancement,[9, 43] and a fingerprint molecule absorption band in ultrasensitive SEIRA spectroscopy.[5-7] Moreover, in the following, we will use SERS to further illustrate this wavelength tuning capability.

Based on the mutual correlation between $\theta_{\text{inc}}$ and the spectral location of the hybrid resonance, the near-field $\langle|\mathbf{E}|^2\rangle$ spectrum can be translated from the wavelength domain (at a specific $\theta_{\text{inc}}$) into the angle domain (at a specific wavelength $\lambda = 637$ nm here, tracked along the horizontal red dotted line in Figure 3A), as seen in Figure 4A. The PCGR-boosted enhancement can be spectrally tuned to match $\lambda = 637$ nm at $\theta_{\text{inc}} = 2.42°$ (also see inset of Figure 4A). When $[\lambda, \theta_{\text{inc}}] = [637$ nm, $8°]$, the hybrid system still provides an ~5× amplification of $\langle|\mathbf{E}|^2\rangle$ relative to that of the solitary AuNR (inset of Figure 4A), although at this moment the PCGR is spectrally away from 637 nm. This broadband, less angle-sensitive (when $\theta_{\text{inc}} \geq 6°$, as seen in Figure 4A) amplification arises from AuNR's coupling to the FP cavity resonance ($Q_{\text{FP}} \approx 19$ for the bare FP cavity) in the TiO$_2$/SiO$_2$/Si layers. When the spectral dip of the FP reflection (that is, the FP cavity resonance[38]) coincides with the AuNR resonance (see Figures 3A and S4), the excited FP standing-wave cavity mode provides constructive optical interference and feedback for the resonant AuNRs on the top surface,[7, 25, 26, 44] enhancing the antenna resonance in a way similar to the PCGR-powered hybrid enhancement, as seen in the near-field distribution in Figure 4B. The ~5× amplification powered by the FP cavity resonance here (relative to the solitary AuNR) is close to amplification values of the similar systems reported previously.[7, 26] In addition, it is important to note that coupling to the PCGR provides an at least one order of magnitude greater amplification of $\langle|\mathbf{E}|^2\rangle$ than that provided by coupling to the FP cavity resonance (Figure 4A).

Next, we experimentally interrogated the numerically predicted synergistic hybrid enhancement with SERS, by probing the two types of cavity-coupling mechanisms that are inherent in the same hybrid architecture, simply by varying $\theta_{inc}$. As seen in Figure 1, the PC slab has a reciprocal lattice vector along the *x* direction only.[35] Therefore, the PCGR is much less sensitive to the incident angle on the *y-z* plane compared to the angle on the *x-z* plane ($\theta_{inc}$).[45] This property inspired us to build a line-focusing fluorescence microscope[46] that not only focuses incident laser on the *y-z* plane in order to increase excitation power density for surface-enhanced spectroscopy, but also retains collimation on the *x-z* plane simultaneously to precisely operate the hybrid system in either the antenna–PCGR or the antenna–FP cavity hybrid state. The schematic diagram of the optical setup is illustrated in Figure 4C (detailed in Supporting Information), describing how $\theta_{inc}$ can be tuned by translationally moving the incidence focus on the back focal plane (BFP) by a displacement $\Delta x$.[38]

The electromagnetic enhancement of the SERS signal scales with the product of the local field enhancements at both the excitation wavelength ($\lambda_{exc}$ = 637 nm here) and a Raman scattering wavelength ($\lambda_{Raman}$).[4] While switching different antenna–cavity coupling mechanisms at $\lambda_{exc}$ (simply by varying $\theta_{inc}$) modifies the near-field intensity for excitation, Raman-scattered photons are emitted at $\lambda_{Raman}$ in identical scenarios. Therefore, measuring SERS signals from *the same* AuNR(s) with excitation at different $\theta_{inc}$ can directly decode their corresponding field enhancements at $\lambda_{exc}$. This method allows us to exclude the influences from differences between samples in configuration, dimension, and surface roughness of the AuNR as well as position, orientation, and number of the adsorbed molecules.[3]

Experimentally measured integrated SERS intensity from surface-bound Rhodamine 6G (R6G) molecules around an individual AuNR is shown in Figure 4D as a function of $\theta_{inc}$, exhibiting a good agreement with the simulated trend in Figure 4A. As also seen in Figure 4E, the hybrid system achieves a strong amplification (>10×) of SERS signal when shifting from the antenna–FP cavity hybrid state (operated at $\theta_{inc} = 8°$) to the antenna–PCGR hybrid state (at $\theta_{inc} = 2.55°$), thereby validating the "upgraded" amplification of $\langle|\mathbf{E}|^2\rangle$ that is predicted in Figure 4A. In addition to SERS from an individual AuNR, collective SERS emission from multiple AuNRs also clearly demonstrates this strong synergistic hybrid enhancement as well (Figure S5), where $x$-oriented AuNRs contribute most of the signal (Figure S3). The sample with randomly oriented AuNRs here also helps validate that the synergistic hybrid enhancement arises from the antenna–cavity coupling, instead of from the inter-antenna interaction.

After showing that the synergistic antenna–PCGR coupling can strongly boost the antenna's near-field intensity, next, using both numerical simulation and an analytical model based on TCMT,[47, 48] we studied this effect as a function of $Q_{PCGR}$. Adjusting the depth $d$ (see Figure 1) varies the radiation loss of the PCGR mode and controls $Q_{PCGR}$ (blue sphere in Figure 5A, at $\theta_{inc} = 0°$), whereas the spectral location of the PCGR is insensitive to changes of $d$. The resulting $Q_{hyb}$ and peak $\langle|\mathbf{E}|^2\rangle$ of the hybrid (with $x$-oriented AuNRs) were obtained using simulation (Figure 5, A and B). When $Q_{PCGR} \leq 300$, $Q_{hyb}$ and $\langle|\mathbf{E}|^2\rangle$ scale with $Q_{PCGR}$. Further increasing $Q_{PCGR}$ to $\geq 10^3$, however, drastically reduces the peak $\langle|\mathbf{E}|^2\rangle$ value, indicating that there exists an optimal $Q_{PCGR}$ for hybrid enhancement.

This dramatic transition can be unraveled by using TCMT. Consider a hybrid system consisting of AuNRs and a PCGR that are on-resonance ($\omega_{PCGR} = \omega_{ant} = \omega_0$, where $\omega_{PCGR} = 2\pi c/\lambda_{PCGR}$ and $\omega_{ant} = 2\pi c/\lambda_{ant}$ are the resonance frequencies of the PCGR and the AuNR, respectively, and $c$ is the speed of light in vacuum). By treating the hybrid system as a resonator, the intensity $\langle |\mathbf{E}|^2 \rangle$ at frequency $\omega$ can be described as (detailed in Supporting Information):

$$\langle |\mathbf{E}|^2 \rangle \propto \frac{\gamma_{rad}}{(\omega - \omega_0)^2 + (\gamma_{rad} + \gamma_{abs})^2}, \qquad [1]$$

where $\gamma_{rad}$ and $\gamma_{abs}$ are the effective loss rates of the hybrid system due to radiation and absorption, respectively. The $Q$-factor of the hybrid resonator is $Q_{hyb} = \omega_0/2(\gamma_{rad} + \gamma_{abs}) = (Q_{rad}^{-1} + Q_{abs}^{-1})^{-1}$, where $Q_{rad} = \omega_0/2\gamma_{rad}$ and $Q_{abs} = \omega_0/2\gamma_{abs}$. Therefore, when the radiative loss of the hybrid is weak ($Q_{rad} \gg Q_{abs}$), then $Q_{hyb} \approx Q_{abs}$. In this regime, a higher $Q_{rad}$ does not further increase $Q_{hyb}$. Moreover, $Q_{abs}$ is proportional to the ratio of the stored optical energy ($U_{em}$) to the dissipation power of the hybrid ($P_{abs}$), and therefore has a weak dependence on $d$ for the same hybrid configuration (see discussion of $Q_{abs}$ in Supporting Information). Accordingly, we extracted the $Q_{abs}$ value by having $Q_{abs} \approx Q_{hyb}$ when $d$ = 5 nm (horizontal green dotted line in Figure 5A), and estimated $(Q_{abs}^{-1} + Q_{PCGR}^{-1})^{-1}$ as a function of $d$ (light blue square in Figure 5A). The excellent agreement between the $Q_{hyb}$ in the simulation and the estimated $(Q_{abs}^{-1} + Q_{PCGR}^{-1})^{-1}$ confirms that the PCGR dominates the radiation loss of the hybrid system ($Q_{rad} \approx Q_{PCGR}$), meaning that radiation loss of the AuNR is strongly suppressed by near-field coupling to the PCGR.

Additionally, TCMT also sheds light on how the peak $\langle |\mathbf{E}|^2 \rangle$ value of the hybrid varies with $Q_{PCGR}$. When the hybrid is excited at resonance ($\omega = \omega_0$), Eq. **1** becomes:

$$\langle |\mathbf{E}|^2 \rangle \propto \frac{Q_{hyb}^2}{Q_{rad}} \approx \frac{Q_{hyb}^2}{Q_{PCGR}}. \quad [2]$$

Using $Q_{hyb}$ and $Q_{PCGR}$ values obtained numerically in Figure 5A, the "radiation engineering" term $Q_{hyb}^2/Q_{PCGR}$ in Eq. 2 shows a trend (red open circle in Figure 5B) in close agreement with the simulated trend of the peak $\langle |\mathbf{E}|^2 \rangle$ value, and therefore disentangles the dependence of the antenna–PCGR cooperativity on $Q_{PCGR}$. Based on the TCMT model, for a given $Q_{abs}$, the highest peak $\langle |\mathbf{E}|^2 \rangle$ value occurs when $Q_{rad}$ ($\approx Q_{PCGR}$) = $Q_{abs}$ = $2Q_{hyb}$ (critical coupling,[16] see Supporting Information for details). In the simulation, $Q_{abs} \approx 304$, and the critical coupling condition can be met when $d \approx 30$–50 nm ($d$ = 32 nm for simulations in Figures 2, 3, and 4).

As shown in Figure 5C, benefiting from the nearly critical coupling ($Q_{PCGR}$ = 222; $Q_{hyb}$ = 133), the hybrid with $d$ = 50 nm demonstrates a 118-fold amplification of $\langle |\mathbf{E}|^2 \rangle$ relative to that of the solitary AuNR (also see Figures 2B and 5B for the similar synergistic enhancement exhibited by the hybrid with $d$ = 32 nm). In contrast, despite an ~45× higher $Q_{PCGR}$ (= 10,168) and an associated strong $\langle |\mathbf{E}|^2 \rangle$ provided by the bare PCGR, as shown in Figure 5D, the hybrid with $d$ = 5 nm only exhibits an 81% lower peak $\langle |\mathbf{E}|^2 \rangle$ value compared to that of the hybrid with $d$ = 50 nm, depicting a low antenna–PCGR cooperativity. Therefore, a judicious selection of the resonant photonic microcavity is crucial for achieving synergistic antenna–cavity hybrid enhancement.

With this understanding, we are in a good position to develop strategies for generating further higher hybrid enhancement, by gaining insights from the TCMT model. Figure 5E shows the two-dimensional contour profile of $Q_{hyb}^2/Q_{PCGR}$ as a function of $Q_{abs}$ and $Q_{PCGR}$ ($\approx Q_{rad}$). For a given $Q_{abs}$, again, the optimal enhancement occurs at critical coupling, $Q_{PCGR} = Q_{abs}$ [for example, see corresponding locations of Figure 5, C and D (blue squares in Figure 5E)]. In the other dimension, for a given $Q_{PCGR}$ [for example, vertical blue dotted line for the $Q_{PCGR}$ value (= 391) of the bare PC slab in the experiment (Figures 2H and S1A)], synergistic hybrid enhancement occurs when $Q_{abs} \geq Q_{PCGR}$, and a lossless hybrid system ($Q_{abs} \to \infty$) has $Q_{hyb}^2/Q_{PCGR} = Q_{PCGR}$, which is 4 times higher than at critical coupling. Therefore, to benefit from high-$Q$ microcavity resonances for synergistic hybrid resonant enhancement, $Q_{abs}$ must be at least comparably high. In comparison to the antenna-enhancement technique using an evanescent diffraction mode supported by a dense array of plasmonic antennas (for example, surface density of gold nanoparticles = 6.25 /μm$^2$ in ref. 23) to capture plane-wave excitation and to provide energy storage and feedback,[6, 23, 24] here the individual antenna is enhanced by the low-loss dielectric PC slab (in addition, the AuNR surface density = 1 /μm$^2$ for the simulations here), which can lead to a lower dissipation density, a larger ratio $U_{em}/P_{abs}$, and therefore a higher $Q_{abs}$ (also see discussion of $Q_{abs}$ in Supporting Information). Absorption losses in our hybrid system mainly come from plasmonic antennas, and a further increase of $Q_{abs}$ can be achieved with engineering of the nanoantenna configuration of the hybrid system, for example, by simply lowering the surface density of plasmonic antennas, or by using low-loss plasmonic materials.[14] A high $Q_{abs}$ value can enable ultrastrong amplification of antenna near fields when a high-$Q$ PCGR is employed. In addition to facilitating ultrasensitive SERS and SEIRA[5] detection, this giant enhancement could also lead to a more efficient plasmonic hot carrier generation[49] than solitary antennas for driving plasmon-induced photodetection and nanochemistry.[8, 50, 51] Another promising alternative method to increase $Q_{abs}$ is to incorporate recent exciting developments in dielectric nanoantennas.[52-56] The "all-dielectric" nanoantenna–microcavity hybrid platform can have ultralow intrinsic absorption losses. Furthermore, while having a comparable near-field enhancement versus its hybrid counterpart equipped with plasmonic antennas, the all-dielectric hybrid system can avoid localized heating of the antenna structures and their immediate surroundings to benefit applications and experiments where local heating is detrimental and unwanted.[57-59] It is also worth comparing the antenna–cavity hybrid strategy with a prior approach that used a ground plane underneath a solitary antenna to control $Q_{rad}$ for critical coupling,[16] where the dielectric spacer thickness of less than quarter-wavelength is too thin to support optical resonance modes[38] and the low $Q_{abs}$ value is dictated by the large dissipative loss of the solitary plasmonic antenna. Here, the cavity-coupling strategy can substantially modify both $Q_{abs}$ and $Q_{rad}$, thereby allowing strong amplification of antenna near fields.

The antenna–PCGR platform offers fabrication advantages and compatibilities, making the synergistic hybrid strategy even more appealing for practical applications. The PC slab here is based on CMOS-compatible materials (Si, $SiO_2$, and $TiO_2$)[60] and processes (see Supporting Information), allowing for scalable and low-cost fabrication of the microcavity structures. The nanoantennas can be chemically synthesized[42, 61, 62] and subsequently placed onto the PC slab surface, just as we have shown in this report, or can be built using top-down lithography. In addition to the intentional control of antenna's orientation by a top-down approach, the nanoantenna–PCGR resonant coupling can also be facilitated by employing a chemically synthesized structure that is polarization-insensitive. In addition to hybrid configurations where antennas are fixed on the substrate surface, the dielectric PCGR substrate can also pair with a movable metallic tip antenna in tip-enhanced Raman spectroscopy (TERS).[63] In this scenario, the coupled tip–PCGR system can resonantly amplify the local field intensity around the tip apex to increase the enhancement and sensitivity in TERS, in contrast to the gap-mode plasmons supported by the tip–metal substrate system used in typical TERS experiments.[63, 64]

In summary, we have demonstrated a highly cooperative hybrid strategy for local field enhancement, by coupling a nanoscale dipolar plasmonic antenna to an on-resonance, modest-$Q$ PCGR microcavity. Predicted with numerical simulation and experimentally validated with SERS, the tunable synergistic interaction in the PCGR-coupled system produces an amplification of the AuNR's local field intensity that is more than one order of magnitude greater than that produced by coupling to the FP cavity. Besides the very simple AuNR, the planar open-access architecture of the PC slab allows easy integration with various advanced plasmonic[1, 3, 7, 63, 65] and dielectric[52-56, 62] nanoantennas for higher performance. Furthermore, our investigations based on TCMT show that, to achieve synergistic hybrid near-field enhancement, it is centrally important to wisely coordinate both $Q_{rad}$ and $Q_{abs}$ of the hybrid system. More generally, we show that increasing $Q_{abs}$ can generate ample room for further hybrid resonant enhancement, thus unleashing the power of the high-$Q$ optical microcavities to benefit a wide range of modern nanoantenna-enhanced applications. As such, the framework presented here provides a technological underpinning to many new possibilities in the future.

**Associated Content**

**Supporting Information.** Materials and methods (numerical simulations, AuNR fabrication, PC slab fabrication, AuNR–microcavity hybrid fabrication; far-field reflection measurements; SERS measurements; estimation of $Q$-factors), modeling a resonant hybrid system (temporal coupled-mode theory, elaboration of $Q_{abs}$, optimal $Q_{rad}$ for a given $Q_{abs}$), optical properties of the bare PC slab, comparisons between the hybrid structures with and without a bottom Si substrate, hybrid enhancement versus AuNR's location and orientation, far- and near-field optical properties of the hybrid versus incident angle, collective SERS emission from multiple AuNRs. Further details are available free of charge at https://pubs.acs.org/doi/suppl/10.1021/acs.nanolett.7b03519.

**Acknowledgements**

This work was financially supported by the National Science Foundation (grant no. 1512043).

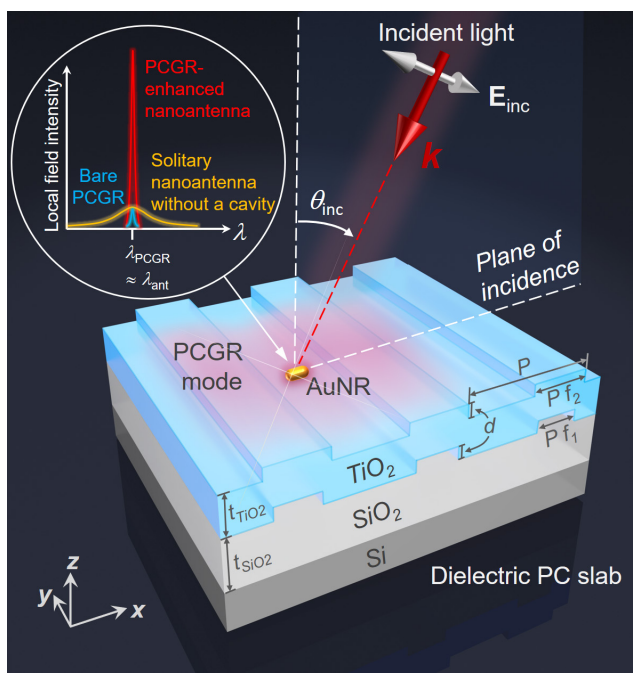

**Figure 1. Conceptual illustration of the synergistic nanoantenna–microcavity hybrid for local field enhancement.** When optical modes of the nanoantenna (AuNR) and the resonant microcavity (PCGR in the dielectric PC slab) overlap in both spatial and spectral domains to form a hybrid supermode, their synergistic interaction can lead to a local field intensity that is orders of magnitude greater than that of the separate entities. Structure parameters of the PC slab: $P$ = 360 nm, $t_{TiO2}$ = 139.6 nm, $t_{SiO2}$ = 725 nm, $d$ = 32 nm, $f_1$ = 33%, and $f_2$ = 45%. The geometry of the AuNR is described in Supporting Information.

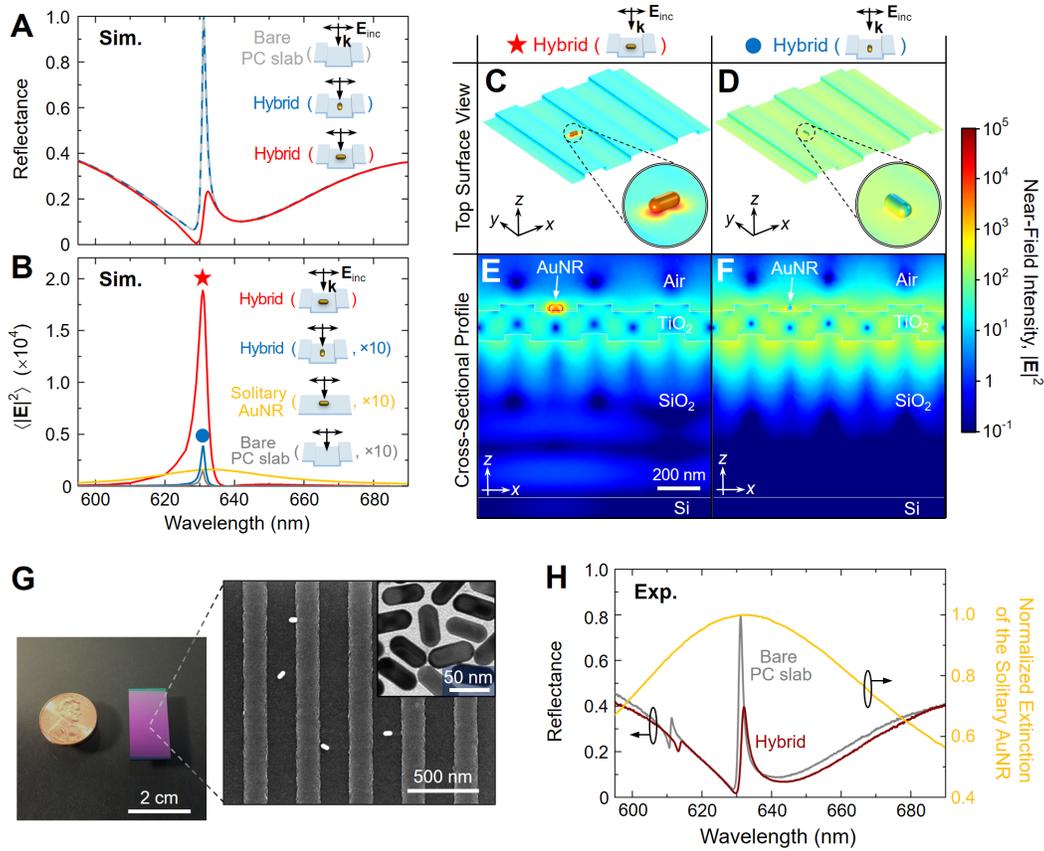

**Figure 2. Far- and near-field optical properties of the AuNR–PCGR hybrid system.** Far-field reflectance (**A**) and average near-field intensity on the AuNR surface $\langle|\mathbf{E}|^2\rangle$ (**B**) computed as a function of wavelength at $\theta_{inc} = 0°$ [red and blue: hybrids with different AuNR orientations; gray: bare PC slab; orange in (B): solitary AuNR on bulk $TiO_2$]. (**C−F**) Simulated surface distributions (**C** and **D**) and cross-sectional slices through the middle of the AuNR (**E** and **F**) of the near-field intensity (on a logarithmic scale) for the hybrid resonances as marked in (B). (**G**) Photograph (left) and top-view scanning electron microscopy (SEM) image (right) of the hybrid, along with transmission electron microscopy (TEM) image of AuNRs (top right inset). (**H**) Experimentally measured reflectance spectra at normal incidence (gray: bare PC slab; wine: hybrid) and extinction spectrum of AuNRs on a flat $TiO_2$-coated glass substrate (orange).

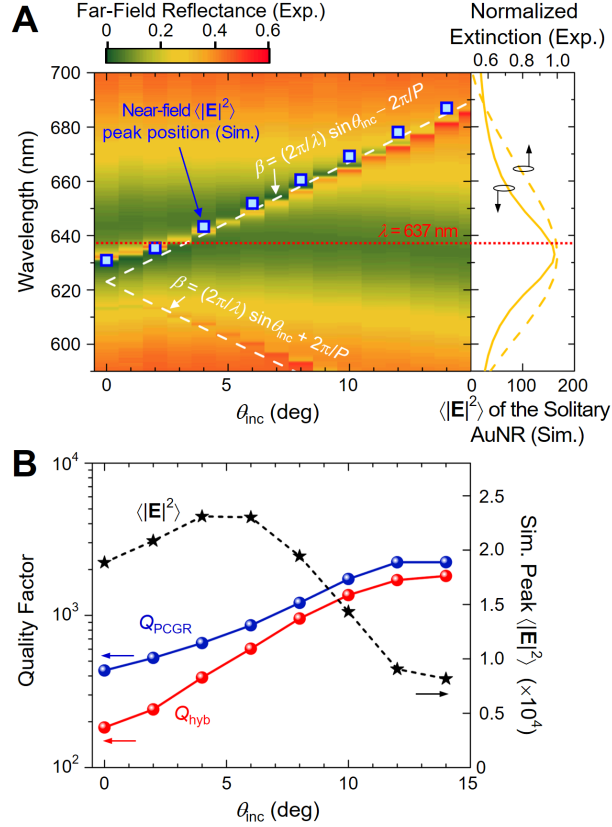

**Figure 3. Optical resonances of the hybrid are spectrally tunable.** (**A**) Experimentally measured far-field reflectance spectrum of the hybrid as a function of $\theta_{\text{inc}}$ (colormap). Spectral locations of the bare $TM_0$ PCGR predicted analytically using the phase-matching condition (white dashed line), together with numerically predicted near-field $\langle|\mathbf{E}|^2\rangle$ peak wavelengths of the hybrid (long-wavelength branch, blue square), are overlaid. Right panel: simulated $\langle|\mathbf{E}|^2\rangle$ spectrum (solid curve) and experimental extinction spectrum (dashed curve) for the solitary AuNR without a cavity. The horizontal dotted red line denotes the 637-nm laser excitation wavelength. (**B**) $Q_{\text{PCGR}}$ (blue sphere), $Q_{\text{hyb}}$ (red sphere), and peak $\langle|\mathbf{E}|^2\rangle$ of the hybrid (black star) evaluated with simulation, as a function of $\theta_{\text{inc}}$.

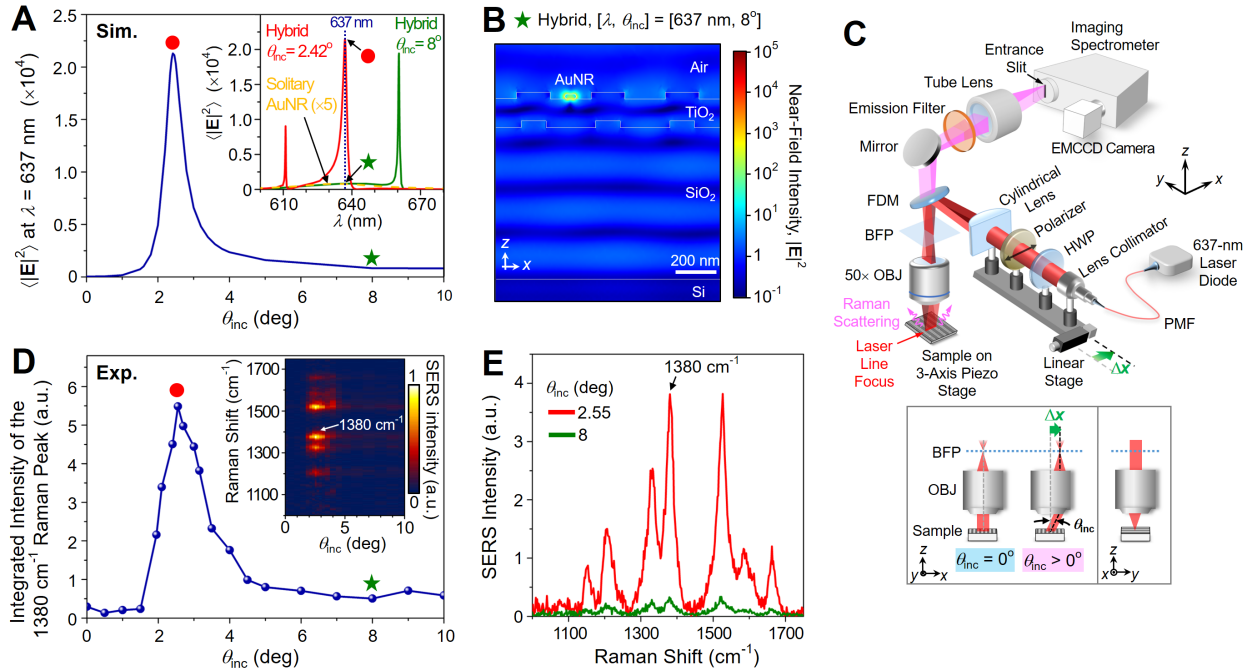

**Figure 4. Predicted synergistic hybrid near-field enhancement is experimentally validated with SERS.** (**A**) Simulated $\langle|\mathbf{E}|^2\rangle$ of the hybrid at $\lambda$ = 637 nm as a function of $\theta_{inc}$. Inset: $\langle|\mathbf{E}|^2\rangle$ spectra of the hybrid (red, $\theta_{inc}$ = 2.42°; green, $\theta_{inc}$ = 8°) and the solitary AuNR on bulk $TiO_2$ (orange dashed). (**B**) Cross-sectional slice of the simulated near-field intensity at [$\lambda$, $\theta_{inc}$] = [637 nm, 8°], corresponding to the antenna–FP cavity hybrid state. (**C**) Schematic illustration of the line-focusing microscopy. HWP, half-wave plate; BFP, back focal plane; FDM, fluorescence dichroic mirror; PMF, polarization-maintaining fiber; OBJ, objective; EMCCD, electron-multiplying charge-coupled device. Bottom inset: illustrations showing angle-tunable collimated incidence on the *x-z* plane while focused light on the *y-z* plane. (**D**) Measured SERS intensity of the 1380 cm$^{-1}$ band (integrated over 1357–1411 cm$^{-1}$) from R6G molecules around an individual AuNR as a function of $\theta_{inc}$. Inset: SERS spectrum as a function of $\theta_{inc}$. (**E**) SERS spectra of the individual AuNR when coupled to two different types of cavity resonances, at $\theta_{inc}$ denoted in (D) (red: coupled to the PCGR; green: coupled to the FP cavity).

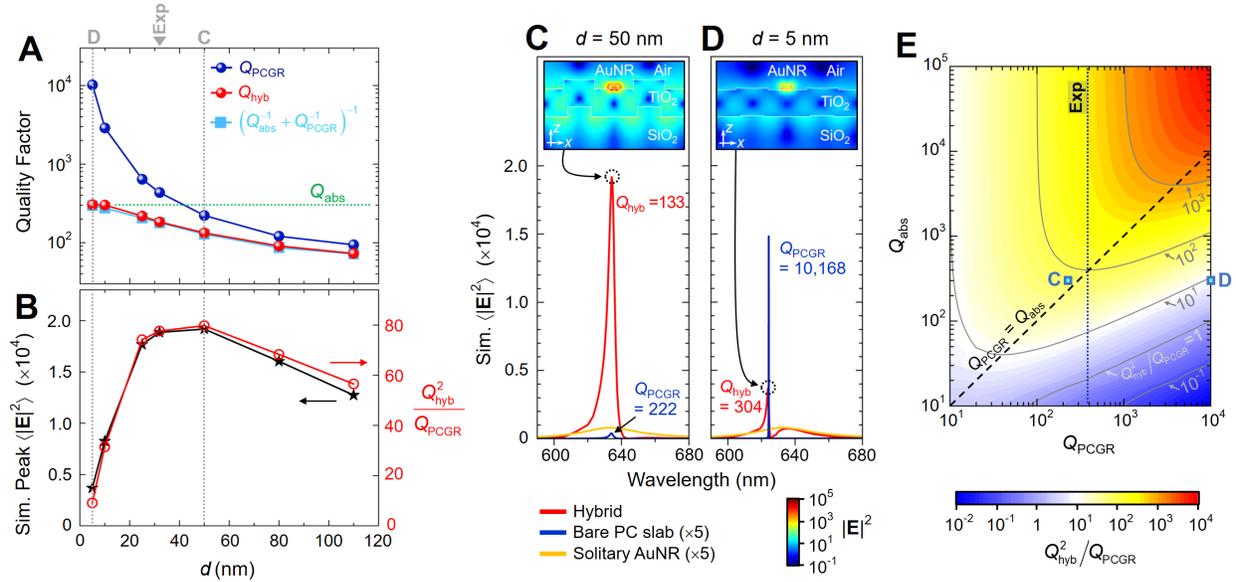

**Figure 5. Synergy of the hybrid requires a right combination of $Q_{PCGR}$ and $Q_{abs}$.** (**A** and **B**) By varying the depth $d$, which controls $Q_{PCGR}$ [blue sphere in (A)], the resulting $Q_{hyb}$ [red sphere in (A)] and peak $\langle|\mathbf{E}|^2\rangle$ value of the hybrid [black star in (B)] were computed with simulation (at $\theta_{inc} = 0°$). Based on the TCMT model, $Q_{abs}$ of the hybrid system was extracted [horizontal green dotted line in (A)], and then $(Q_{abs}^{-1} + Q_{PCGR}^{-1})^{-1}$ was evaluated [light blue square in (A)] to compare with simulated $Q_{hyb}$. The term $Q_{hyb}^2/Q_{PCGR}$ in Eq. 2 was evaluated as a function of $d$ [red open circle in (B)] to explain the simulated trend of the peak $\langle|\mathbf{E}|^2\rangle$ value in (B). (**C** and **D**) Simulated $\langle|\mathbf{E}|^2\rangle$ spectra of the hybrid, the bare PC slab, and the solitary AuNR for two representative $d$ values [also denoted in (A) and (B), vertical gray dotted line], along with corresponding spatial near-field profiles at resonance (sliced through the middle of the AuNR), exhibiting distinctly different antenna–cavity cooperativities. (**E**) Contour map of $Q_{hyb}^2/Q_{PCGR}$ as a function of $Q_{PCGR}$ and $Q_{abs}$, where $Q_{hyb} = (Q_{abs}^{-1} + Q_{PCGR}^{-1})^{-1}$. The critical coupling condition (black dashed line), $Q_{PCGR}$ in the experiment (with $d = 32$ nm, vertical blue dotted line), and corresponding locations of (C) and (D) (blue square) are denoted.

**TOC Graphic**

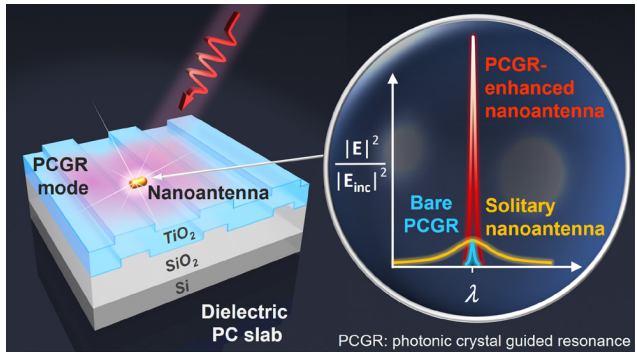